\documentclass[10pt, conference, letterpaper]{IEEEtran}
\usepackage{cite}
\usepackage{graphicx,epstopdf,url}
\usepackage{amsmath}
\usepackage{amssymb}
\usepackage{amsfonts}
\usepackage{graphicx}
\usepackage{verbatim}
\usepackage{textcomp}

\usepackage{booktabs}
\usepackage[utf8]{inputenc}

\usepackage{xcolor}
\usepackage{soul}
\usepackage{tabularx}
\usepackage{makecell}
\usepackage{wrapfig}
\usepackage{footnote}
\usepackage{bbm}
\usepackage{algorithm}
\usepackage{algpseudocode}
\usepackage{subcaption}
\usepackage{indentfirst}
\usepackage[all]{xy}
\usepackage{bbm}
\newcommand{\our}{DIMEC-DC}

\def\BibTeX{{\rm B\kern-.05em{\sc i\kern-.025em b}\kern-.08em
    T\kern-.1667em\lower.7ex\hbox{E}\kern-.125emX}}
\begin{document}

\title{Distributed Inference on Mobile Edge and Cloud: A Data-Cartography based Clustering Approach}


\author{\IEEEauthorblockN{Divya Jyoti Bajpai and Manjesh Kumar Hanawal}
\IEEEauthorblockA{\textit{MLioNS Lab, Dept. of IEOR, IIT Bombay} \\
Mumbai, Maharashtra-400076, India\\
\{divyajyoti.bajpai, mhanawal\}@iitb.ac.in}
}

\maketitle

\begin{abstract}
The large size of DNNs poses a significant challenge for deployment on devices with limited resources, such as mobile, edge, and IoT platforms. To address this issue, a distributed inference framework can be utilized. In this framework, a small-scale DNN (initial layers) is deployed on mobile devices, a larger version on edge devices, and the full DNN on the cloud. Samples with low complexity (easy) can be processed on mobile, those with moderate complexity (medium) on edge devices, and high complexity (hard) samples on the cloud. Given that the complexity of each sample is unknown in advance, the crucial question in distributed inference is determining the sample complexity for appropriate DNN processing. We introduce a novel method named \our{}, which leverages the Data Cartography approach initially proposed for enhancing DNN generalization. By employing data cartography, we assess sample complexity. \our{} aims to boost accuracy while considering the offloading costs from mobile to edge/cloud. Our experimental results on GLUE datasets, covering a variety of NLP tasks, indicate that our approach significantly lowers inference costs by more than 43\% while maintaining a minimal accuracy drop of less than 0.5\% compared to performing all inferences on the cloud \footnote{The source code is available at https://anonymous.4open.science/r/DIMEC-1B04}.
\end{abstract}


\section{Introduction}

In recent years, the size of Deep Neural Networks (DNNs) has expanded considerably, leading to exceptional performance, particularly in Natural Language Processing (NLP) tasks \cite{han2022survey, wang2019glue}. However, this increased scale demands extensive computational resources, posing challenges for deployment on resource-limited platforms such as mobile and edge devices. To mitigate these challenges, several approaches have been developed, including model pruning \cite{zhu2017prune, michel2019sixteen}, weight quantization \cite{zhang2020ternarybert, kim2021bert}, knowledge distillation \cite{sanh2019distilbert, jiao2019tinybert}, early exits \cite{xin2020deebert, zhou2020bert, bajpai2024ceebert}, and cloud offloading \cite{matsubara2022split}.

Approaches such as model pruning, weight quantization, and knowledge distillation aim to reduce model size through various techniques, often leading to a notable decrease in model accuracy. These methods typically compress the model to fit within the memory limits of mobile devices, which can compromise the backbone's effectiveness. Many large DNN models have smaller versions tailored for resource-constrained environments \cite{devlin2018bert, carreira2023revolutionizing}. For instance, the BERT \cite{devlin2018bert} model has smaller variants like DistilBERT \cite{sanh2019distilbert} and TinyBERT \cite{jiao2019tinybert} for mobile deployment, BERT-base for edge deployment, and BERT-Large and Extra-Large for cloud deployment. Although these smaller models are computationally efficient and fit in mobile devices, they suffer a significant loss in accuracy.

Mobile and edge devices frequently struggle with performing inference on large models due to limitations in space and memory. Cloud offloading takes advantage of high-capacity services and extensive computational resources, enabling the deployment of full-scale DNNs for inference tasks. Nevertheless, offloading samples to the cloud introduces extra costs due to the physical separation from mobile terminals.

\begin{figure}
\centering
\includegraphics[scale=0.5]{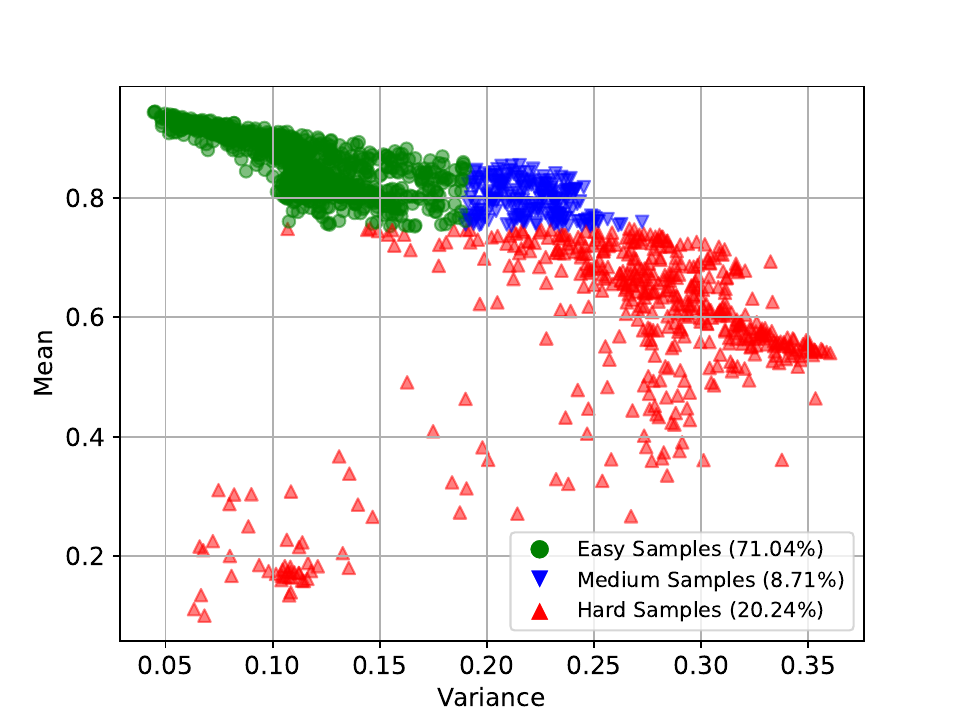}
\caption{The figure shows the clustering of samples based on the confidence and variance obtained during training across epochs, where green samples are easy samples, blue samples are moderate, and red are hard samples.}
\label{fig: Train_accuracy}
\end{figure}

In real-world applications, the input tasks contain a mix of easy and hard samples, which do not all require the same level of computational effort. For instance, Figure \ref{fig: Train_accuracy} displays a plot of the mean confidence and variance of samples across epochs. Here, confidence and variance are defined as the mean and variance of the ground-truth label probabilities across epochs for the SST-2 dataset using the BERT-base model \cite{devlin2018bert}. The green samples, which have high confidence and low variance, are easily learned by the model. Conversely, the blue samples, which exhibit high confidence but also high variance, indicate that the model's confidence in the ground-truth label fluctuates. The challenging samples with high variance are those where the model is less confident; however, a larger model can learn to correctly classify these samples. Finally, the most difficult samples are those with low variance but where the model is consistently incorrect; these are often out-of-distribution samples, making accurate prediction challenging even with a full-fledged DNN.

This analysis demonstrates that not all samples require the same amount of computation; easier samples can achieve correct inference locally, even with smaller models deployed on resource-constrained devices, thus significantly reducing cost and time complexity. However, harder samples necessitate the use of a full-fledged DNN for accurate inference, which is not feasible on resource-constrained devices.

To tackle this issue, we employ a distributed inference strategy: deploying the initial layers of the DNN on the mobile device, a more extensive model with additional layers on the edge device, and the complete model on the cloud. Considering the diverse complexity of real-world samples, it is beneficial to leverage mobile, edge, and cloud resources according to the complexity of incoming samples. Since the complexity of these samples is not predetermined, the challenge lies in accurately assessing them. We propose a method to determine the complexity of each sample based on {\it data cartography} \cite{swayamdipta2020dataset}. It enables the decision of whether the sample should be processed on the mobile device, edge device, or cloud. The inference location decision is crucial as it balances accuracy and various costs; while processing more samples on the cloud can achieve higher accuracy, it incurs greater offloading costs. Conversely, processing more samples on the mobile or edge device can compromise accuracy while keeping the cost low.

Data Cartography is a technique designed to enhance the generalization capability of DNNs. It identifies the most ambiguous samples within a large dataset, pinpointing those that are most challenging for the model to classify. This approach leverages the prediction confidence across multiple training epochs to gauge the complexity of samples. Samples with high confidence and low variance are deemed `easy', those with low confidence are considered `hard' irrespective of variance, and samples with high confidence and high variance are seen as `ambiguous', indicating the model's fluctuating confidence and confusion.

\begin{figure*}
\centering
\includegraphics[scale=0.59]{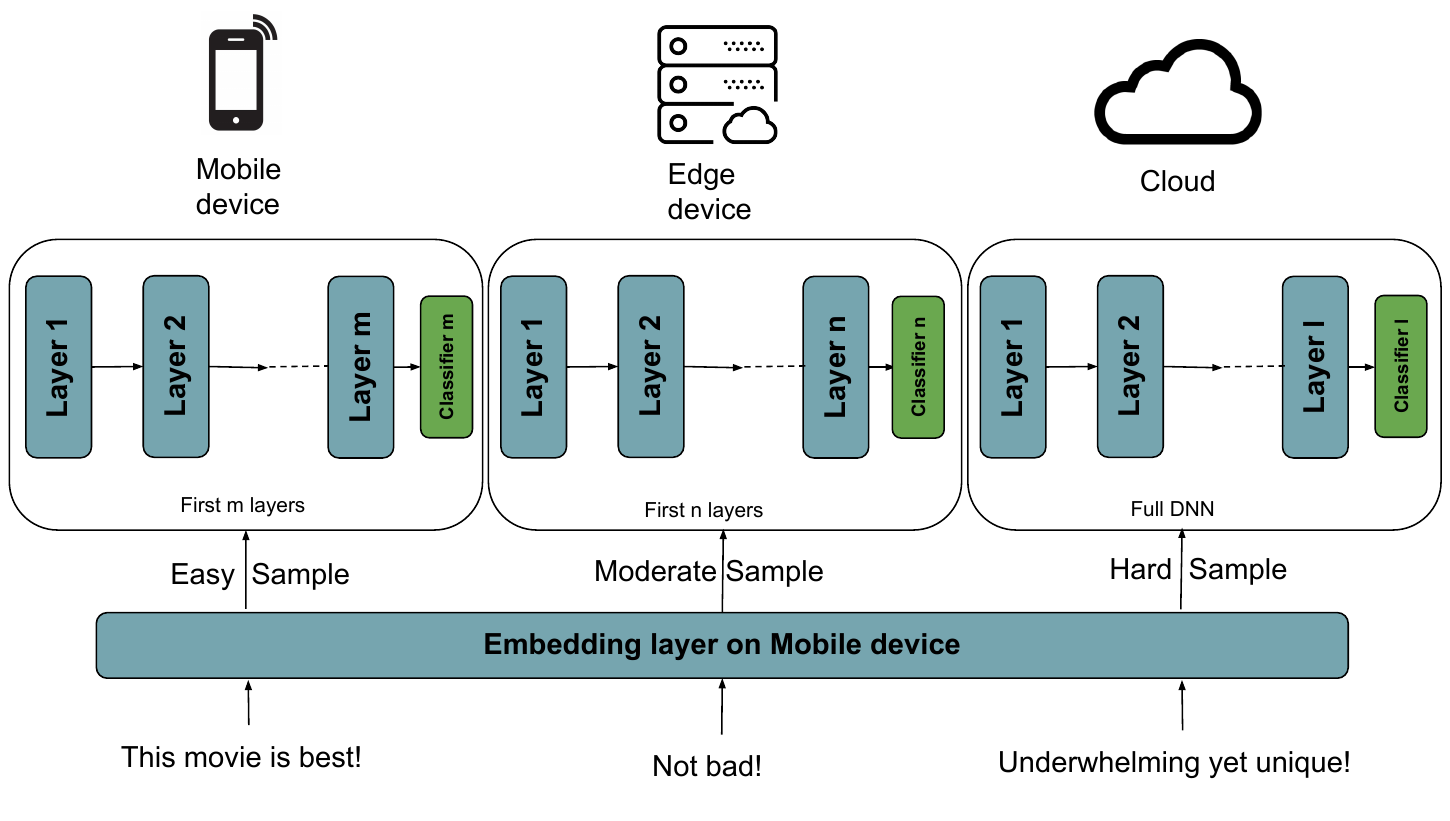}
\caption{In this figure, three types of reviews are input to the mobile device. It passes through the embedding layer on the mobile device where it decides about the complexity of the sample. The DNN is divided into three parts: 1) First $m$ layers are deployed on the mobile device and easy samples are then inferred on the mobile device. 2) First $n$ layers are deployed on the edge device and the sample that is more complex than it can not be inferred on the mobile is inferred at the edge. 3) Finally, a fully-fledged DNN is deployed on the cloud and the sample is offloaded only if it falls in the hardest pool of samples, i.e., both mobile and edge do not have sufficient layers of the DNN to correctly predict it.}
\label{fig: main}
\end{figure*}

Our method enhances resource efficiency across mobile, edge, and cloud environments through distributed inference, leveraging the data cartography technique. We deploy three versions of the DNN: the initial layers on the mobile device, a larger portion on the edge device, and the complete model on the cloud. The allocation of layers to each device is determined based on the available resources on the mobile and edge devices, as detailed in Section \ref{sec:layer}. Each device includes a classifier at its final layer, enabling inference at each device. Figure \ref{fig: main} illustrates our inference process, where easy samples are handled by the mobile device, moderately complex samples are offloaded to the edge, and only the most challenging samples are offloaded to the cloud for processing.

To determine sample complexity during inference, we categorize samples into easy, moderate, and hard pools by assessing their confidence and variance over various training epochs. To create these pools, we initially utilize a smaller version (initial layers) of the backbone model intended for deployment on a mobile device. The model's weights from each training epoch are utilized to compute the confidence and variance of the samples on the validation dataset. Based on these metrics, we classify the samples' complexity and append their word embeddings to the corresponding easy, medium, or hard pool. For example, if a sample is classified as easy, its word embedding is added to the easy pool. During inference, we utilize these pre-established pools to evaluate the complexity of incoming samples and assign them to an appropriate device. It is important to note that only the embedding layer is required on the mobile device to determine the samples' complexity.

In our method, named \our{}: \underline{D}istributed \underline{I}nference on \underline{M}obile, \underline{E}dge, and \underline{C}loud: A \underline{D}ata-\underline{C}artography based Clustering Approach, we leverage distributed inference to allocate computational resources according to the complexity of each sample. This method dynamically determines the computational needs of an incoming sample on the fly, avoiding the necessity for samples to be processed by the mobile device first. This alleviates the load on individual devices and addresses the challenge of running large models on mobile devices. Additionally, our method effectively balances the accuracy-efficiency trade-off, significantly enhancing efficiency while maintaining accuracy comparable to that of the final layer. This strategy ensures efficient and accurate processing by dynamically deciding whether to process samples locally, at the edge, or offload them to the cloud based on their complexity, thus optimizing both processing and communication costs.

We utilize the BERT-base/large \cite{devlin2018bert} model as our backbone due to its well-established performance and efficiency, making it an ideal candidate for evaluating our approach. This choice allows us to assess our method's effectiveness across various NLP tasks, as described in Section \ref{sec: experiments}. Our experiments, which include sentiment classification, entailment classification, and natural language inference, reveal that \our{} is resilient to different cost structures, effectively accommodating devices with diverse processing capabilities and communication technologies such as 3G, 4G, 5G, and Wi-Fi. Notably, our method achieves a substantial cost reduction ($>43\%$) while incurring only a slight decrease in accuracy ($<0.5\%$) compared to scenarios where all samples are processed at cloud.

Our main contributions are summarized as follows:

\begin{itemize}
\item We develop a mechanism based data cartography to decide if a sample can be inferred at DNN available on mobile, edge or cloud.
\item Our approach demonstrates resilience to variations in cost structures, maintaining accuracy even when mobile and edge devices or communication networks are altered.
\item The negligible reduction in accuracy is due to our method preserving the backbone's optimality, as it does not involve any parameter reduction from the model.
\item Through empirical evaluation, we show that our method effectively reduces costs and, in some cases, improves performance compared to previous baseline approaches.
\end{itemize}

\section{Related Works}
In this section, we discuss the existing methods for inference on resource-constrained devices.

Neurosurgeon, as introduced in \cite{kang2017neurosurgeon}, explores the strategies for optimizing the splitting of DNNs based on cost considerations associated with selecting a specific splitting layer. In a similar vein, BottleNet \cite{eshratifar2019bottlenet} and Bottlefit \cite{matsubara2022bottlefit} incorporate a bottleneck mechanism within split computing.  This approach involves deploying part of the DNN on a mobile device to encode the input into a compact representation, which is then processed further in the cloud. Various training strategies have been proposed for training the encoder on the edge device within this setup such as BottleNet++ \cite{shao2020bottlenet++} and \cite{hu2020fast} employs cross-entropy-based training approaches in the context of split computing, Matsubara \cite{matsubara2019distilled} performs knowledge distillation-based training, CDE \cite{sbai2021cut} and Yao \cite{yao2020deep} perform reconstruction-based training 
and Matsubara \cite{matsubara2021neural} perform head-network distillation training method to effectively encode the input to offload efficiently.

The cloud offloading methods utilize split computing where part of the DNN is deployed on mobile and the remaining on the cloud where the sample is encoded into smaller bits so that offloading cost can be reduced. But in this setup, all the samples are required to be offloaded to the cloud irrespective of sample complexity. Also, some computation is performed on the mobile device for each sample hence the method adds to both mobile processing cost as well as offloading cost.

AdaEE \cite{pacheco2021distorted} employs a combination of Early Exit DNNs \cite{zhou2020bert, xin2020deebert, bajpai2024ceebert} and DNN partitioning to facilitate offloading data from mobile devices to the cloud using early exit DNNs. LEE (Learning Early Exit) \cite{ju2021learning}, DEE (Dynamic Early Exit) \cite{ju2021dynamic} and UEE-UCB \cite{hanawal2022unsupervised} leverage the multi-armed bandit (MAB) framework to determine optimal exit points up to which the DNN can be processed on the mobile device, while SplitEE \cite{bajpai2023splitee} and I-SplitEE \cite{bajpai2023splitee, bajpai2024splitee} utilizes the MAB \cite{ML02_UCB1_Auer} setup to get the optimal splitting points under domain change scenarios in edge-cloud co-inference setup. LEE and DEE are specifically designed for efficient edge device inference, particularly in cases of service disruptions, employing utility functions that require ground-truth labels.

The above-mentioned methods utilize cloud offloading with early exits to perform dynamic inference on mobile or edge devices. Although these methods have the option to perform inference on mobile devices based on sample complexity, these methods decide on the sample complexity after processing the sample through a mobile device, hence all samples add to the processing cost.

Our approach differs from previous works in various aspects: 1) We are the first ones to leverage the data cartography method for clustering to perform distributed inference. 2) During inference, our method does not require processing the sample on the mobile device and then offloading instead it decides on the fly, which device could be appropriate based on sample complexity. 3) Only hard samples are offloaded to the cloud and easy and medium ones are inferred locally on mobile or edge with smaller cost requirements.

\section{Data cartography}
In this section, we provide a concise overview of the data cartography technique, which is an approach designed to identify the most beneficial data from a large corpus. This technique addresses the question of how datasets can be categorized based on their contribution to achieving optimal performance both within and outside of distribution. By leveraging training dynamics—specifically, the behavior of the model throughout the training process—data cartography categorizes datasets into three categories: easy, hard, and ambiguous. The classification is based on the confidence and variance values observed across training epochs for each sample, as illustrated in Figure \ref{fig: Train_accuracy}. The primary objective is to identify a subset of the dataset that enhances the model's generalization capabilities.

\subsection{Training dynamics}

Consider a training dataset of size $N$, $\mathcal{D} = \{(x_i, y_i^{})\}_{i=1}^{N}$, where the $i$th instance consists of an observation $x_i$ and its ground-truth label $y_i^{*}$. The model used for training outputs a probability distribution over labels given an observation. A stochastic-gradient descent-based optimization procedure is employed, with training instances randomly ordered at each epoch, across $E$ epochs.
The training dynamics of instance $i$ are defined as statistics calculated across the $E$ epochs. These measures serve as the coordinates in the data map. The first measure captures how confidently the learner assigns the true label to the observation, based on its probability distribution. Confidence is defined as the mean model probability of the true label ($y_i^{*}$) across epochs:
\begin{equation}\label{eq: mean}
    \hat{\mu}_i = \frac{1}{E}\sum_{e = 1}^{E} \mathcal{P}(y_i^*|x_i, \theta_e),
\end{equation}
where $\mathcal{P}$ denotes the model probability and $\theta_e$ represents the parameters at the end of the $e$th epoch.

The second coordinate for the data map is variability, which measures the spread of $\mathcal{P}(y_i^*|x_i, \theta_e)$ across epochs using variance (The paper \cite{swayamdipta2020dataset} originally uses standard deviation, but we use the variance directly):

\begin{equation}\label{eq: variance}
    \hat{\sigma}_i = {\frac{\sum_{e = 1}^{E} \left(\mathcal{P}(y_i^{*}|x_i, \theta_e)-\hat{\mu}_i\right)^2}{E}} 
\end{equation}
The variability also depends on the ground-truth label $y_i^{*}$. Instances for which the model predicts the same label consistently (whether accurately or not) will have low variability, whereas instances, where the model is indecisive across training, will exhibit high variability.

Most samples fall into the high-confidence and low-variability region of the map (Figure \ref{fig: Train_accuracy}, top-left). The model consistently predicts these instances correctly with high confidence, labeling them as easy-to-learn. A smaller group is characterized by low variability and low confidence (Figure \ref{fig: Train_accuracy}, bottom-left). These samples are rarely predicted correctly during training and are thus referred to as hard-to-learn. The third group includes ambiguous examples with high variability (Figure \ref{fig: Train_accuracy}, right-hand side); the model's confidence in these instances fluctuates, indicating indecision. These are referred to as ambiguous samples.

Following this classification of training data, the model is trained on the top 33\% of samples from each group. It has been demonstrated that training on ambiguous samples yields the best generalization performance on out-of-distribution tasks.


\section{Problem setup}
In this section, we will discuss the adaptation of Data Cartography techniques for distributed inference. Note that we have a different objective of assessing the complexity of an unseen sample hence instead of training, we use the validation dataset to create the clusters. The confidence in the validation set instances is slightly lower than that of the training set as it is unseen to the backbone during training. We also need to find the thresholds that can perfectly classify the data such that the accuracy is not degraded.

We start with a pre-trained language model such as BERT. We assume that during inference, the first $m$ layers will be deployed on the mobile, the first $n$ layers on the edge and the full BERT model on the cloud. Hence during training, we attach classifiers at the $m$th and $n$th layers that map the hidden representations of the BERT model to class probabilities. This makes the model capable of performing inference on any device. Also, this corresponds to the case similar to a smaller model such as TinyBERT (4 layers) on mobile, DistilBERT (6 layers) on the edge and full-fledged BERT-base (12 layers) on the cloud.


\subsection{Training exit classifiers}
Recall that we used $\mathcal{D}$ to represent the distribution of the dataset with a label class $\mathcal{C}$ used for backbone fine-tuning. Let us assume that there are $l$ layers in the backbone. For any input sample, $(x,y^*)\sim \mathcal{D}$ and for any classifier $h$, the loss can be computed as:

\begin{equation}
    \mathcal{L}_h(\theta) = \mathcal{L}_{CE}(f_h(x, \theta), y^*) 
\end{equation}

Here, $f_h(x, \theta)$ is the output of the classifier at the $h$th layer, where $\theta$ denotes the set of learnable parameters, and $\mathcal{L}_{CE}$ is the cross-entropy loss. We learn the classifiers of all the devices simultaneously hence the overall loss function could be written as $\mathcal{L} = \mathcal{L}_m+\mathcal{L}_e+\mathcal{L}_c$ where $\mathcal{L}_m, \mathcal{L}_e$ and $\mathcal{L}_c$ are the losses of the classifiers at the mobile, edge and cloud. Also, let $\hat{\mathcal{P}}_h(c)$ denote the estimated probability class $c\in \mathcal{C}$.

\subsection{Preparation of dataset pool}
During the model training phase, we perform the training for $E$ epochs. After every epoch during training, the model weights are saved. Then we obtain the values of $\hat{\mu}_i$ and $\hat{\sigma}_i$ using equation \ref{eq: mean} and \ref{eq: variance} on the validation dataset with the parameters saved after every epoch. In our case, the probability is calculated at the $m$th layer i.e., the mean value is now:

\begin{equation}
    \hat{\mu}_i = \frac{1}{E}\sum_{e = 1}^{E} \hat{\mathcal{P}}_m(y_i^*|x_i, \theta_e)
\end{equation}

Similarly $\hat{\sigma}_i$ is computed,
after computing $\hat{\mu}_i$ and $\hat{\sigma}_i$, we define two thresholds for $\alpha$ and $\beta$ to decide the complexity of the sample. To create the pool, if confidence exceeds the threshold $\alpha$ and variance is less than $\beta$, we classify that as an easy sample. If the confidence is less than the threshold $\alpha$ and irrespective of variance we add the sample to the hard pool of samples. Finally, if a sample has confidence greater than $\alpha$ and variance greater than $\beta$, then the sample is added to the medium pool of samples.
This procedure is given in the Algorithm \ref{alg:pool}. Note that $embed(x)$ is the output of the embedding layer.

\begin{algorithm}
\caption{Pool creation}\label{alg:pool}
\begin{algorithmic}[1]
\State {\textbf{Input:} $x$ and threshold $\alpha$ and $\beta$}
\State \textbf{Initialize} $P_e = [ ], P_m [ ], P_h = [ ]$

\For{$e \gets 1 \text{ to } E$}
\State Calculate $\mu_i = \hat{\mathcal{P}}_m(y^*|x, \theta_e)$
\EndFor
\State Calculate $\hat{\mu}_i$, $\hat{\sigma}_i$ using equation \ref{eq: mean} and \ref{eq: variance}.

\If{$\hat{\mu}_i\geq\alpha$  and $\hat{\sigma}_i<\beta$}
\State $P_e.append(embed(x))$
\ElsIf{$\hat{\mu}_i<\alpha$ and $\hat{\sigma}_i\geq\beta$}
\State $P_m.append(embed(x))$
\Else
\State $P_h.append(embed(x))$
\EndIf
\State \textbf{Return: $P_e, P_m, P_h$}

\end{algorithmic}
\end{algorithm}

\subsection{Choice of threshold $\alpha$ and $\beta$}
The threshold $\alpha$ and $\beta$ are a crucial part of our pool creation as the easiness and hardness of the pool depend on the values of these parameters. It also models the size of the pools as a smaller value of $\alpha$ can increase the number of easy samples and can even classify some hard samples as easy while a larger value can add the easy sample to the harder pool. Both these scenarios affect the model performance as the former case affects the accuracy of the model and the latter case affects the cost.  
Hence it is very crucial to set the threshold properly. 

In that line, we first define the different types of costs that we consider, 1) Processing cost is the cost to process the sample through one layer of the DNN in the mobile and edge denoted as $\lambda_m$ and $\lambda_e$ respectively. 2) Offloading cost from mobile to edge and mobile to cloud denoted as $o_e$ and $o_c$ respectively. We also assume that there is a constant cost $\gamma$ charged by the cloud platform for each sample. We also define $C_h$ as the confidence score in the estimate at the $h$th classifier i.e., $C_h:=\max_{c\in\mathcal{C}}\hat{\mathcal{P}}_h(c)$, it is the maximum of the output probability distribution at the $h$th classifier. Note that $C_h$ is different from $\hat{\mu}$ as this does not have any ground-truth label requirements.
To choose the threshold $\alpha$ and $\beta$, we define a reward function as:

\begin{equation}\label{eq:reward}
    r(\alpha, \beta) = \left\{
        \begin{array}{ll}
            C_m-\lambda_m m & \textit{if} \quad \hat{\mu}_i \geq \alpha \text{ and } \hat{\sigma}_i\leq \beta\\
             C_n- \lambda_e n-o_e & \textit{if} \quad \hat{\mu}_i \geq \alpha \text{ and } \hat{\sigma}_i> \beta\\
              C_l- o_c - \gamma& \textit{ otherwise }
        \end{array}
    \right.
\end{equation}

The reward function could be interpreted as, if an observation $x$ falls in the easy pool of sample i.e., $x\in P_e$ then it is to be inferred at the mobile device i.e., at the $m$th layer hence the reward will be the confidence $C_m$ at the $m$th layer subtracted by the cost of processing the sample on the mobile device. Similarly, if the observation falls in a medium pool of samples i.e., $x\in P_e$, then it will be inferred at the edge i.e., at the $n$th layer then the reward will be the same as that of the mobile device with an additional cost of offloading and a reduced processing cost. Finally, if the sample falls in the hard pool of samples i.e., $x\in P_h$, then the sample will be inferred at the cloud i.e., at the final layer of the BERT, then the reward will be the inference at the final layer subtracted by the cost of the cloud platform and offloading cost. The expected reward function could be written as:

\begin{multline}
\mathbb{E}[r(\alpha, \beta)] = \mathbb{E}[C_m-\lambda_m i|\text{mob. inference}] P[\text{mob. inference}]\\ + \mathbb{E}[C_n-\lambda_e-o_e|\text{edge inference}] P[\text{edge inference}]\\
\mathbb{E}[C_l-\gamma-o_c|\text{cloud inference}] P[\text{cloud inference}]
\end{multline}

Now the objective is to maximize $\mathbb{E}[r(\alpha, \beta)]$ and could be expressed as $\max_{(\alpha, \beta)\in S_1\times S_2}\mathbb{E}[r(\alpha, \beta)]$ where the set $S_1$ and $S_2$ are the possible choices for the $\alpha$ and $\beta$ values respectively. $P[\text{mob. inference}]$, $P[\text{edge inference}]$ and $P[\text{cloud inference}]$  is the probability that the sample will be inferred at mobile, edge and cloud respectively and depend on the value of $\alpha$. Note that we maximize this reward function on the validation dataset as we already have the pools created for the validation split of the dataset.


\subsection{Post-Deployment Inference}
\textbf{Fixed:} After storing the values $P_e$, $P_m$ and $P_h$ consisting of embeddings of easy, moderate and hard samples respectively. We calculate the average of these pools and name it as $P_e^a$, $P_m^a$ and $P_h^a$ respectively. The sample can be classified as easy, moderate or hard using the average values as in K-means clustering algorithm, as a sample arrives, the distance of the incoming sample is calculated from $P_e^a$, $P_m^a$ and $P_h^a$ and classifies the sample as easy, moderate or hard based on the minimum distance of the sample from the mean values of different pools. After this the easy samples are inferred locally at the mobile device incurring only processing cost, moderate samples are offloaded directly to the edge without any computation on mobile incurring small offloading cost and processing cost and the hard samples are directly offloaded to the cloud with higher offloading cost as well as cost charged by the cloud platform.

\textbf{Adaptive:} In fixed inference, the pools are created using the validation dataset, however during test time there might be a shift in the dataset distribution. For such cases, we dynamically update the pool averages such that the distribution shift can be properly captured. In this setup, as the sample arrives, it is classified as easy, moderate or hard in a similar way but this time the average is recalculated based on the incoming sample's complexity. For instance, if a sample is easy, then the value the average $P_e^a$ is recalculated and updated. In this manner, the shift in the test dataset is captured and the trade-off of accuracy-cost is not affected. 

\begin{algorithm}
\caption{Dynamic Inference}\label{alg:inference}
\begin{algorithmic}[1]
\State {\textbf{Input:} Test sample $x$, $P_e^a, P_m^a and P_h^a$}
\State $n_e, n_m, n_h =$ number of easy, medium and hard samples. 
\State $x_e \gets embed(x)$
\State Calculate distance from all pool averages $d(x_e, \_)$
\State $d_{min}(x) \gets \min\{d(x_e, P_e^a), d(x_e, P_m^a), d(x_e, P_h^a)\}$
\If{$d_{min} = d(x_e, P_e^a)$}

\State Infer the sample locally on mobile.
\State $P_e^a\gets\frac{n_e.P_e^a+x_e}{n_e+1}$, $n_e\gets n_e+1$
\ElsIf{$d_{min} = d(x_e, P_m^a)$}
\State Offload the sample to the edge and infer.
\State $P_m^a\gets\frac{n_m.P_m^a+x_e}{n_m+1}$, $n_m\gets n_m+1$
\Else
\State Offload the sample to the cloud and infer.
\State $P_h^a\gets\frac{n_h.P_h^a+x_e}{n_h+1}$, $n_h\gets n_h+1$
\EndIf

\end{algorithmic}
\end{algorithm}

In Algorithm \ref{alg:inference}, we only show the adaptive version of our method. To obtain the fixed version from given algorithm, lines 8, 11, 14 need not be executed. However, we prefer the adaptive inference as it performs better as shown in the experiments and also comes with almost negligible computational cost.

\subsection{Analysis of Layer distribution}\label{sec:layer}
We assume that the mobile device contains DNN's first $m$ layers and the edge has DNN's first $n$ layers where $1\leq m\leq n\leq l$ where the cases $m=1$ is when there is no mobile device, $m=n$ means there is either no mobile or edge device. If $n = l$, it means there is no edge device. 
We discuss the impact of the values of $m$ and $n$. These values are important as they model the overall cost and are user-defined. They are used to decide the quantity of workload on different devices, i.e., mobile, edge or cloud. A higher value of $m$ means more layers are deployed on the mobile device and the processing cost e.g. battery depletion will be high, however since more layers are in the mobile device there will be a lower chance of a sample being offloaded reducing the latency cost. If the value of $n$ is high, then there will be fewer samples being offloaded to the cloud reducing the latency costs, however, it will increase load on the edge device. If both $m$ and $n$ are kept small then since less number of layers will inferred earlier, more samples will be offloaded to the cloud increasing the offloading cost and the charges of the cloud platform.

\section{Experiments}\label{sec: experiments}

\begin{table*}[]
\centering
\begin{tabular}{ccccccccccccc}
\hline
Model/Data & \multicolumn{2}{c}{SST-2}    & \multicolumn{2}{c}{CoLA}   & \multicolumn{2}{c}{MNLI}     & \multicolumn{2}{c}{MRPC}     & \multicolumn{2}{c}{QNLI}     & \multicolumn{2}{c}{QQP}    \\ \hline
           & Acc           & Cost         & Acc         & Cost         & Acc           & Cost         & Acc           & Cost         & Acc           & Cost         & Acc         & Cost         \\ \hline
BERT       & 93.5          & 1.00            & 58.3        & 1.00            & 84.5          & 1.00            & 89.2          & 1.00            & 92.5          & 1.00            & 72.1        & 1.00            \\
Random     & 89.5          & -27\%          & 55.7        & -31\%          & 79.9          & -46\%          & 86.5          & -39\%          & 89.6          & -49\%          & 69.4        & -32\%          \\
DeeBERT & 91.0            & -23\%          & 56.5        & -25\%          & 82.1          & -31\%          & 87.6          & -42\%          & 90.2          & -36\%          & 70.0          & -28\%          \\
AdaEE      & 92.1          & -36\%          & 56.9        & -40\%          & 82.8          & -42\%          & 88.1          & -51\%          & 91.4          & -41\%          & 70.8        & -30\%          \\
I-SplitEE  & 92.4          & -45\%          & 57.3        & -39\%          & 83.6          & -48\%          & 88.5          & -58\%          & 91.9          & -57\%          & 71.3        & -39\%          \\ \hline
Ours-F     & 93.0          & -55\%          & 57.3        & -45\%          & 83.9          & -52\%          & 88.4          & \textbf{-65}\% & 91.7          & -61\%          & 71.1        & -36\%          \\
Ours-D     & \textbf{93.2} & \textbf{-61}\% & \textbf{57.9} & \textbf{-50}\% & \textbf{84.1} & \textbf{-59}\% & \textbf{88.7} & -62\%          & \textbf{92.1} & \textbf{-64}\% & \textbf{72.2} & \textbf{-43}\% \\ \hline
\end{tabular}
\caption{Main results of the paper on the BERT backbone. The baseline cost is considered as the original BERT model deployed on the cloud.}
\label{tab: results}
\end{table*}

In this section, we provide all the experimental details of the paper and experimentally validate our method. 

\subsection{Dataset}
We used the GLUE \cite{wang2019glue} datasets for the evaluation of our method. We evaluate our method on three types of tasks viz. sentiment classification, entailment classification and natural language inference. The datasets used are: 

1) MRPC: Microsoft Research Paraphrase Corpus is a semantic equivalence classification dataset containing sentence pairs extracted from online news sources. 2) QQP: Quora Question Pairs is also a semantic equivalence classification dataset but the sentence pairs are extracted from the community question-answering website Quora. 3) SST-2: Stanford-Sentiment Treebank is a sentiment classification dataset. 4) CoLA: Corpus of Linguistic Acceptability with a task of linguistic acceptability of a sentence.
5) QNLI: Question-answering natural language inference is a dataset with a labelling task indicating whether the answer logically entails the question's premise. 6) MNLI: Multi-Genre Natural Language Inference also contains sentence pairs as premise and hypothesis, the task is to classify them as entailment, contradiction or neutral.

\begin{figure*}
    \centering
    \begin{subfigure}{0.319\textwidth}
        \includegraphics[width=\textwidth]{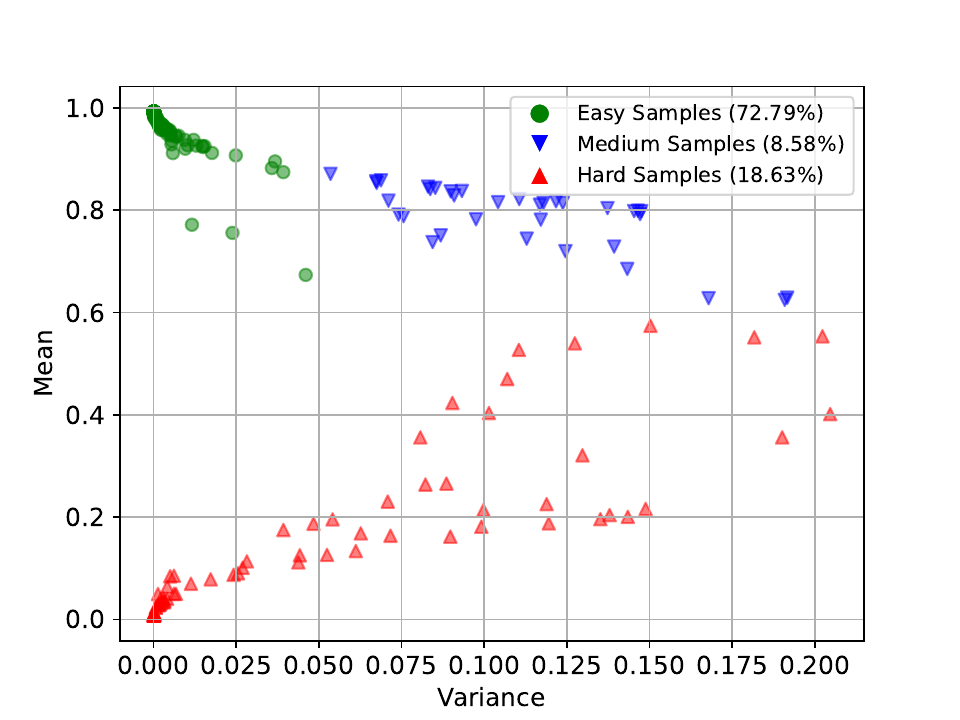}
        \caption{MRPC dataset}
        \label{fig:clust_mrpc_val}
    \end{subfigure}
    \begin{subfigure}{0.319\textwidth}
        \includegraphics[width=\textwidth]{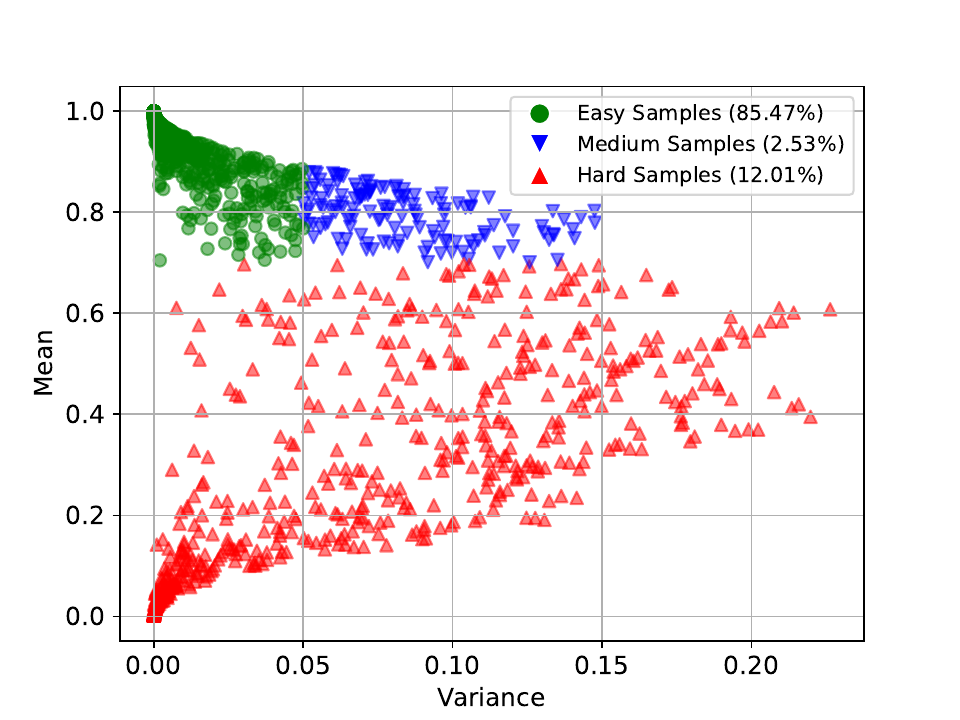}
        \caption{QNLI dataset}
        \label{fig:clust_qnli_val}
    \end{subfigure}
    \begin{subfigure}{0.319\textwidth}
        \includegraphics[width=\textwidth]{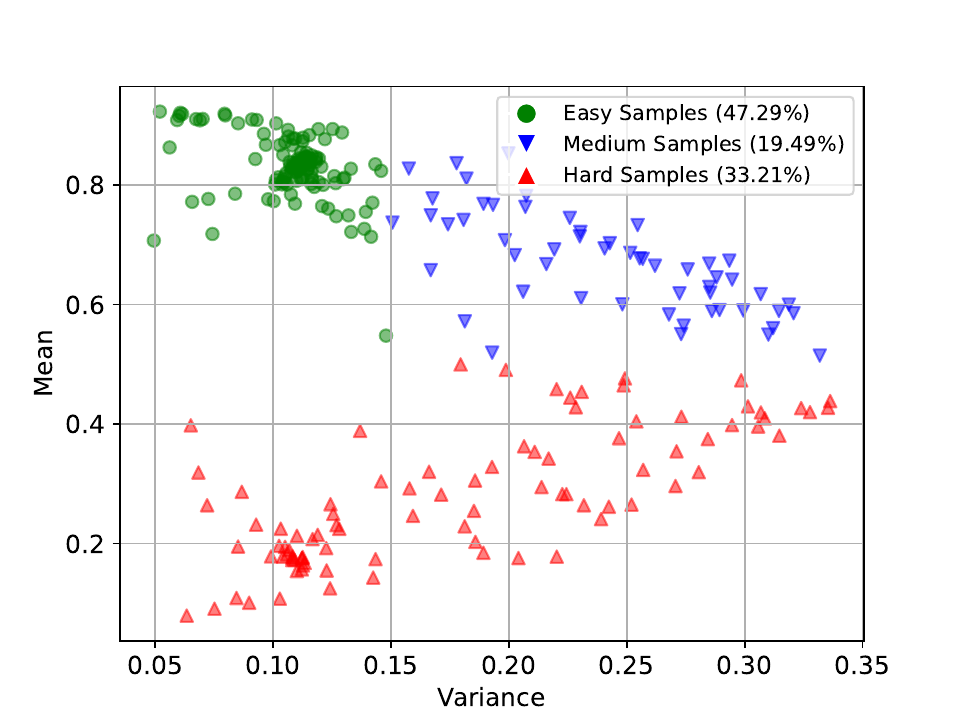}
        \caption{SST-2 dataset}
        \label{fig:clust_sst_val}
    \end{subfigure}
    \begin{subfigure}{0.319\textwidth}
        \includegraphics[width=\textwidth]{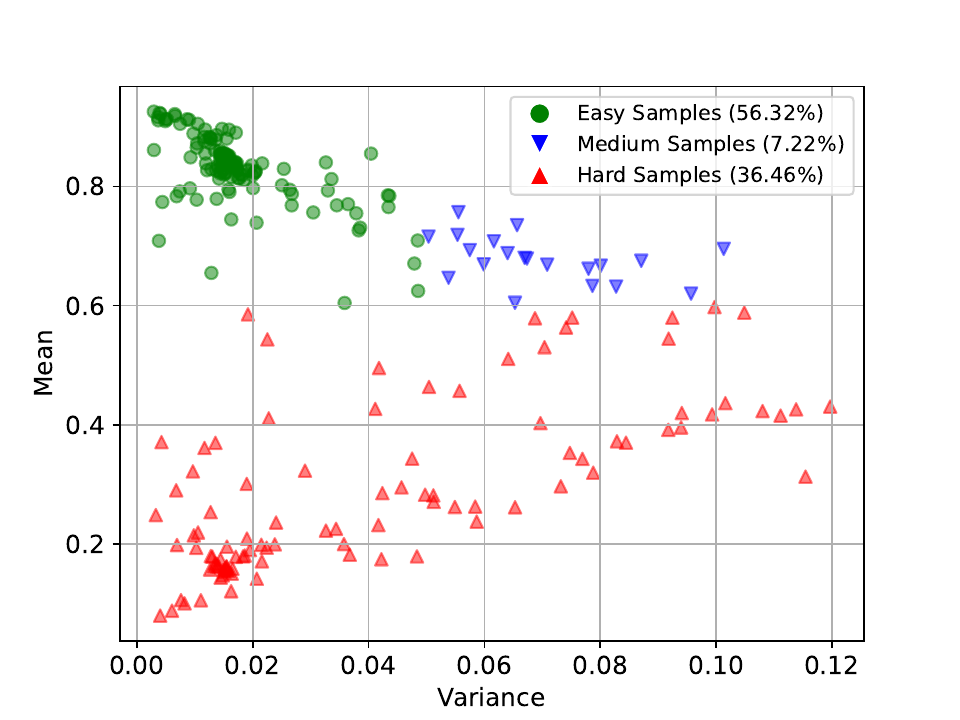}
        \caption{RTE dataset.}
        \label{fig:clust_rte_val}
    \end{subfigure}
    \begin{subfigure}{0.319\textwidth}
        \includegraphics[width=\textwidth]{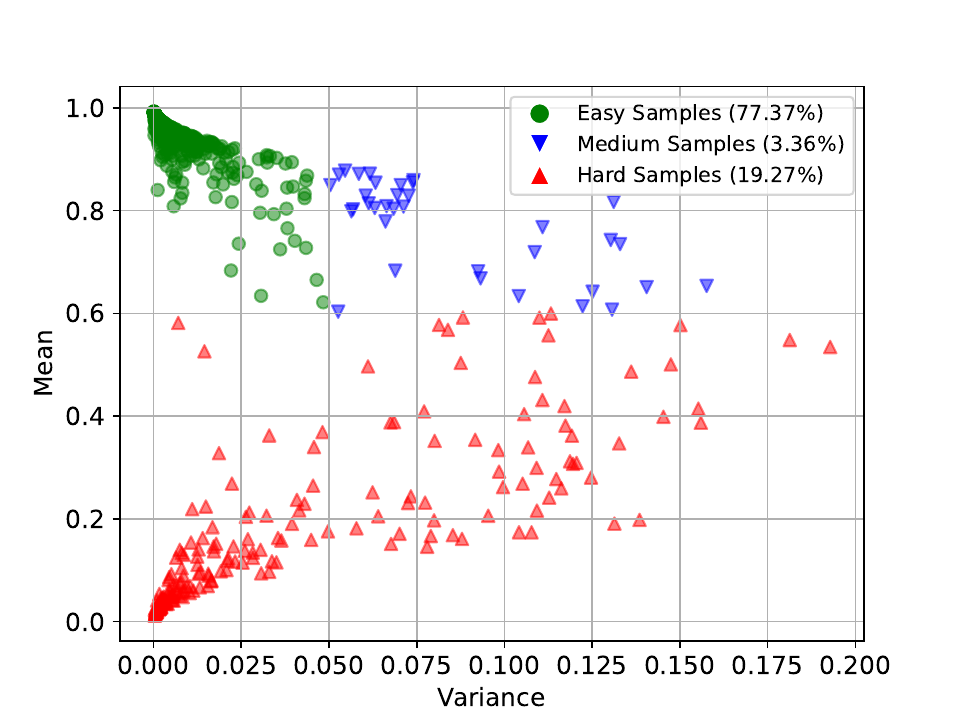}
        \caption{CoLA dataset}
        \label{fig:clust_cola_val}
    \end{subfigure}
    \caption{The figure shows the clustering of samples based on the confidence and variance on the validation split of the datasets, the figures show the initial proportion of samples in different clusters.}
    \label{fig:global}
\end{figure*}

\subsection{Baselines}
We compare the model against various baselines that are detailed below:

\textbf{1) BERT model:} In this baseline, we report the results of the original BERT backbone. We assume that the BERT model is deployed on the mobile device and only processing cost is incurred. This is the main baseline for us.

\textbf{2) Random: } In this baseline, the incoming sample is randomly assigned to one of the given devices i.e. mobile, edge or the cloud. This is created to show that our assignment based on the pooling of samples makes a significant difference.

\textbf{3) DeeBERT:} is the baseline where we assume that the model is deployed completely on the cloud device but with attached exits i.e., it lowers the charge by the cloud as then only partial resource is used for each sample. This baseline shows that splitting the model also helps due to the presence of hard samples.

\textbf{4) AdaEE:} This method is an adaptive method that uses multi-armed bandits to learn the optimal threshold to decide offloading in an edge-cloud co-inference setup. 

\textbf{5) I-SplitEE:} This method learns the optimal splitting layer based on the accuracy-cost trade-off in an online setup. The method uses multi-armed bandits to learn the optimal layer in an edge-cloud co-inference setup where the test dataset contains distortions. 

\textbf{6) Ours-F:} is our method that uses a fixed pool average and does not update it during inference.

\textbf{7) Ours-D:} is our method that dynamically updates the pool averages and covers any domain shift occurring during inference.

We use the same hyperparameters for all the baselines as given in their respective codebases. The cost for all the baselines is calculated using our cost structure which is very similar to most of the previous methods.
There are three key phases in our experimental setup.

\subsection{Training the backbone}
To evaluate our method, we use the widely accepted BERT-base/large model. We add a linear output layer after the final layer on the mobile and, after the final layer on the edge to act as a classifier that maps the intermediate outputs of the backbone to class probabilities. We split the dataset into three parts: 80\% for training, 10\% for validation and 10\% for test. We train the backbone using the train split. We run the model for $5$ epochs. We also perform a grid search over a batch size of $\{8,16,32\}$ and learning rates of \{1e-5, 2e-5, 3e-5, 4e-5, 5e-5\} with Adam \cite{kingma2014adam} optimizer. We apply an early stopping mechanism and select the model with the best performance on the validation set.

\subsection{Pool creation and cost}
We create the pool using the validation split of the dataset. The values of $m$ and $n$ are chosen using the cost structure and we choose $m = 4$, $n=6$ and $m = 6$, $n= 12$ for BERT-base and large respectively, for the BERT-base model it maps to TinyBERT (4 layers) on mobile, DistilBERT (6 layers) on edge and BERT-base on the cloud while on BERT-large (24 layers), it corresponds to DistilBERT on mobile, BERT-base on edge and BERT-large on the cloud. The set of thresholds for confidence and variance are chosen from the set $S_{\alpha} = \{0.55, 0.6, 0.65, 0.7, 0.8\}$ and $S_{\beta} = \{0.05, 0.08, 0.11, 0.14, 0.17\}$. Hence, the search space for $(\alpha, \beta)$ is $S_{\alpha}\times S_{\beta}$. The values of $\alpha$ and $\beta$ are chosen based on the pair that maximizes the reward in equation \ref{eq:reward} on the validation split of the dataset.


Recall, that we have denoted the processing cost for the mobile device as $\lambda_1$ and the processing cost for the edge device as $\lambda_2$, $o_1$ as the offloading cost for mobile to edge and the offloading cost for mobile to cloud as $o_2$. We also assume the cost charged by the cloud platform as $\mu$.

We convert all the costs in terms of the smallest unit. Without loss of generality, we assume the smallest cost as the processing cost of the edge device, we assume $\lambda_2 = \lambda$, $\lambda_1 = (3/2)\lambda$, $o_1 = (5/2)\lambda$ and finally $o_2 = 3\lambda$ to show the results but in the ablation studies, we experiment by varying these costs (see section \ref{sec:cost_vars}). For results obtained in Table \ref{tab: results} we have fixed these values, however, in section \ref{sec:cost_vars}, we show that how the cost changes affect the accuracy and overall cost, this ablation study shows that our model can perform better with any mobile, edge and cloud device as the performance does not degrades with changes in cost. The choice of the cost values is user-specific and processing cost could be chosen as the mobile and edge device computational power and offloading costs depend on the communication networks. 


\subsection{Inference}
During inference, we use a batch size of $1$ as data arrives sequentially. As a sample arrives, the word embedding of a sample is obtained on the mobile device. Then the distance of the word embedding of the sample is calculated against pool averages and the sample is assigned to the closest pool average. If the closest pool is the easy pool, then the sample is inferred on the mobile device. If the closest is the moderate pool, then the sample is offloaded to the edge device. Else, the sample is offloaded to the cloud. 

All the experiments were conducted on NVIDIA RTX 2070 GPU with an average runtime of $\sim$ 3 hours and a maximum runtime of $\sim$ 10 hours for the MNLI dataset.

\section{Results}
In Table \ref{tab: results} and \ref{tab:res_large}, we show the main results of our paper, our method outperforms all the existing baselines both in terms of cost as well as accuracy for both BERT-base and large models. The reduction in cost is larger for the BERT-large model which is intuitive as the large variant is more overparameterized.
\begin{table}[ht]
\centering
\begin{tabular}{ccccccc}
\hline
Model/Data & \multicolumn{2}{c}{CoLA}     & \multicolumn{2}{c}{MRPC}     & \multicolumn{2}{c}{QNLI}     \\ \hline
           & Acc           & Cost         & Acc           & Cost         & Acc           & Cost         \\ \hline
BERT       & 59.5          & 1.00         & 90.1          & 1.00         & 93.1          & 1.00         \\
Random     & 55.9          & -45\%          & 85.2          & -59\%          & 90.4          & -51\%          \\
DeeBERT & 56.2          & -32\%          & 88.2          & -48\%          & 91.2          & -55\%          \\
AdaEE      & 58.1          & -54\%          & 89.2          & -57\%          & 91.7          & -62\%          \\
I-SplitEE  & 58.4          & -57\%          & 88.9          & -58\%          & 92.4          & -68\%          \\ \hline
Ours-F     & 58.5         & -68\%          & 89.7          & -63\%          & 92.1          & -70\%          \\
Ours-D     & \textbf{58.9} & \textbf{-71\%} & \textbf{90.1} & \textbf{-58\%} & \textbf{92.5} & \textbf{-75\%} \\ \hline
\end{tabular}
\caption{Results on the BERT-large variant}
\label{tab:res_large}
\end{table}
\begin{figure*}
    \centering
    \begin{subfigure}{0.32\textwidth}
        \includegraphics[width=\textwidth]{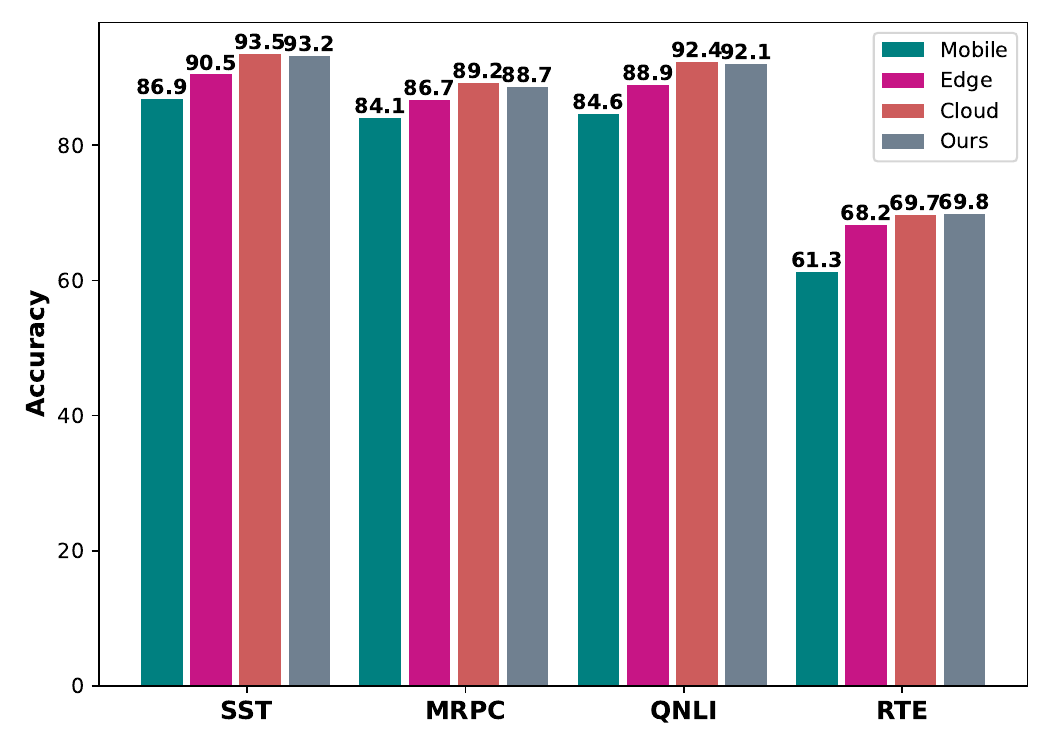}
        \caption{Accuracy of Mobile, Edge and Cloud}
        \label{fig:accuracy}
    \end{subfigure}
    \begin{subfigure}{0.32\textwidth}
        \includegraphics[width=\textwidth]{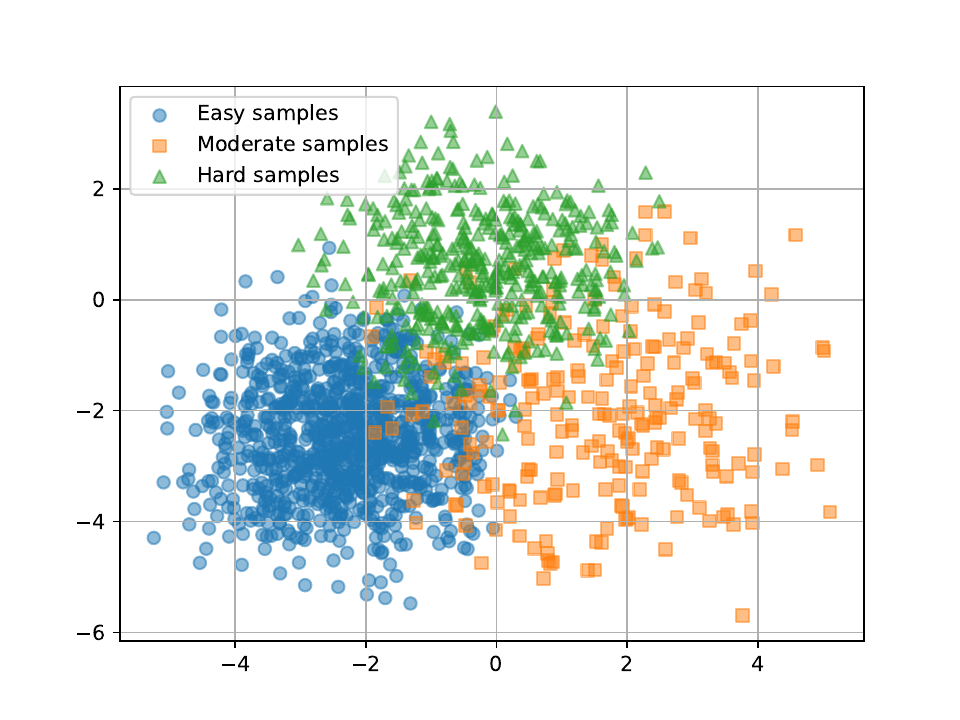}
        \caption{The t-SNE visualization for QNLI}
        \label{fig:cluster_qnli}
    \end{subfigure}
    \begin{subfigure}{0.32\textwidth}
        \includegraphics[width=\textwidth]{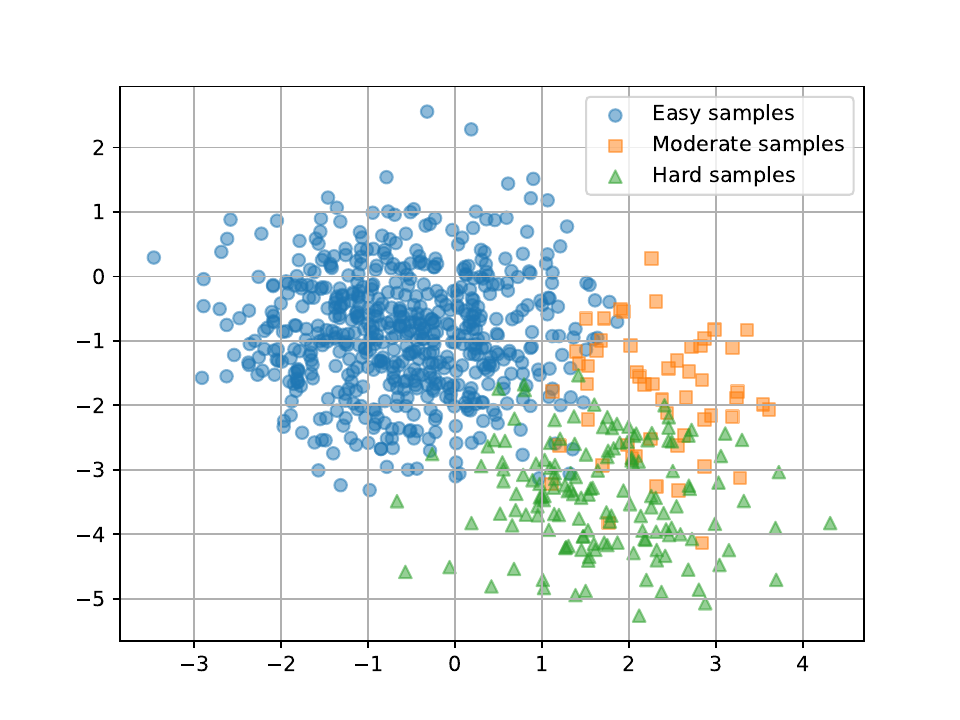}
        \caption{The t-SNE visualization of SST-2.}
        \label{fig:cluster_sst}
    \end{subfigure}
    \caption{The figure shows the accuracy of the individual devices i.e., mobile, edge and cloud. Figure on centre and right: The t-SNE visualization of the word embeddings of the easy, moderate and hard pool created for the QNLI and SST-2 datasets.}
    \label{fig:global_new}
\end{figure*}

\begin{figure*}
    \centering
    \includegraphics[scale = 0.35]{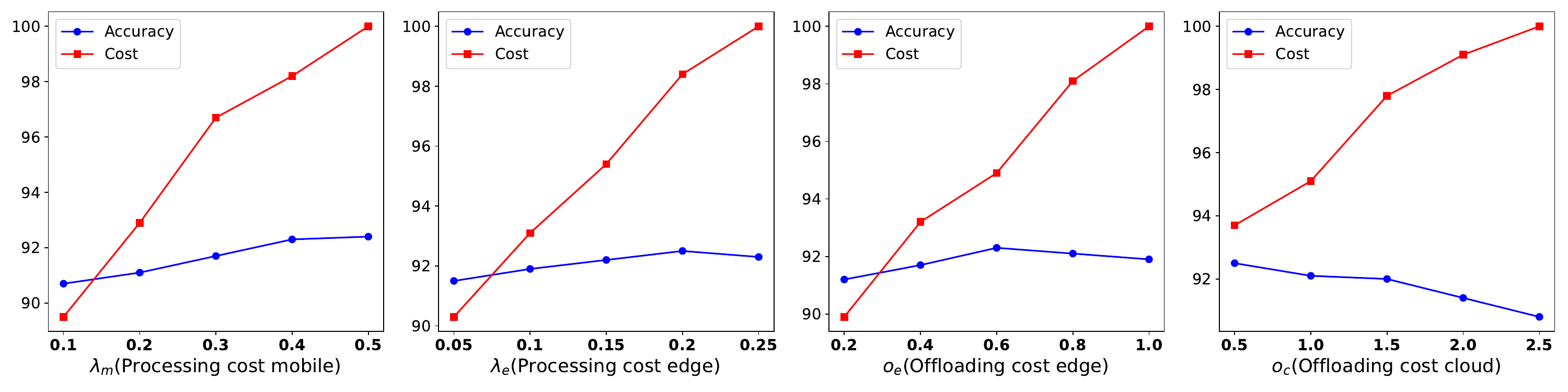}
    \caption{The changes in accuracy and percentage change in cost values when one of the costs is varied while keeping others at a constant value.}
    \label{fig:cost_variations}
\end{figure*}
The BERT model has a higher cost as all the samples are required to pass through the final layer and there is no inference happening on mobile or edge devices. Due to this the accuracy of this model is comparable to our method. In the random assignment of samples to any of the devices, the loss in accuracy is due to the fact that sometimes even the hard samples are assigned to the mobile device, while the increase in cost is due to the fact that easy samples are sometimes assigned to the cloud. The vanilla early exit method DeeBERT gets a lower accuracy as the threshold chosen for exiting is not chosen based on any optimization algorithm but only through a validation set, also this model is deployed on the cloud with lesser charges as the sample might not require to pass-through the complete backbone. 

AdaEE also has lower accuracy as the method mostly works better under domain change scenarios, but in our case the domain shift is minor, it simply reduces to an early exit model with dynamic learning of threshold. Due to this dynamic learning of threshold, it outperforms vanilla early exiting. Finally, the I-SplitEE model also has lower accuracy again due to the case that, it works better in domain shift scenarios. In terms of cost, these models are higher as they require all the samples to be processed on the mobile device before offloading.

Our method outperforms all the baselines, the higher accuracy comes from the appropriate assignment of the sample to various devices and a smaller cost as compared to other methods since all the complexity of the sample is decided based on the word embedding that does not require much processing on mobile reducing the processing cost to a larger extent while in other methods, this cost is very high as it is required for all the samples.

Also, note that for the QQP dataset our method even outperforms the vanilla BERT inference, this is because of the overthinking issue during inference. This issue occurs when an easy sample is passed through the complete backbone leading to the extraction of irrelevant features which in turn results in a wrong prediction as pointed out in \cite{zhou2020bert}.



\section{Ablation Study and Discussion}
In this section, we perform ablation studies and also discuss the choice of layers using the computational powers of mobile and edge and offloading costs.
\subsection{Individual device inference}
We stated that our method uses a distributed inference method between mobile, edge and cloud. In Figure \ref{fig:accuracy}, we show the effect on accuracy when all the samples are inferred on one of the given devices. It means that instead of distributing the inference, performing the inference on a single device. We plot the accuracies of the individual devices and our model. Since the cloud contains the full-fledged DNN, it has the highest accuracy; however, our method sometimes outperforms the full-fledged DNN due to the overthinking issue in DNNs. 

\subsection{Cost Variations}\label{sec:cost_vars}
In Figure \ref{fig:cost_variations}, we show the variation in accuracy and cost if we alter one of the given costs. In the left figure in Figure \ref{fig:cost_variations}, we alter $\lambda_m$ i.e., the processing cost of mobile device, while keeping other costs constant. The accuracy is not affected in this case as we are increasing the processing cost that forces more samples to offload to the edge and cloud having more layers hence accuracy slightly improves. Similarly, if the processing cost $\lambda_e$ is increased, accuracy again slightly improves. As we increase the offloading cost of the cloud $o_c$, we observe that there is a drop in accuracy. This is expected as higher offloading costs for the cloud will set lower thresholds such that most of the samples are inferred locally or at the edge and do not offload to the cloud which in turn lowers the accuracy. Note that in this setup, other costs are kept constant. Also, our method is robust to changes in different types of costs i.e., the loss in accuracy is minimal when cost values vary.

\subsection{Cluster visualization}
In figure \ref{fig:global}, we visualize the classification made based on the complexity of the samples on the validation dataset, these represent the percentage of samples in each cluster, after this set, these clusters are updated if the dynamic inference is performed, otherwise, these remain the same. Recall that these clusters are made after solving equation \ref{eq:reward}. Also observe that all these curves have a bell-shaped structure where the easy samples possess high-confidence and low-variance values, while the hard samples possess low confidence and low variance, the samples with high variance are mostly the ones where the model is confused. One key observation is that for slightly harder datasets such as RTE and MRPC, where even the vanilla BERT is not able to gain higher accuracy, it has classified more samples as hard while for slightly easier datasets, it has classified more datasets as easy. This is due to the reward structure as it maximizes the confidence while keeping the confidence lower.

\subsection{t-SNE Visualizations}
 In Figure \ref{fig:cluster_qnli}, \ref{fig:cluster_sst}, we show what the pool looks like after creation. In this, we show the t-SNE (t-distributed Stochastic Neighbor Embedding) plots of the word embeddings that are used to visualize higher dimensional data. The t-SNE brings the high-dimensional data into two-dimensional space. We can observe from the plots that there is a distinction between the clusters where only the moderate samples have a higher overlap with the other two. Also, observe that the spread of the moderate samples is high, this could be due to the clustering, where we have considered moderate samples as those which have higher variance.

\section{Conclusion}
We address the inference of large DNNs on mobile devices using the complexity of the input samples. We propose a method that utilises data cartography to decide the complexity of samples. It minimizes the cost of inference by assigning the appropriate amount of resources required to infer the incoming sample between mobile, edge and cloud. If the task is easy and require less computation then most of the samples are inferred locally while if the task is hard, then most of the samples are offloaded maintaining accuracy. Our method is robust to changes in cost values i.e., various mobile and edge devices. Experiments on various NLP tasks show the significance of our work where the drop in accuracy is $(<0.3\%)$ while reducing the cost $(>43\%)$ as compared to final layer on the cloud. 

\bibliographystyle{IEEEtran}
\bibliography{mab}

\end{document}